\documentclass[prb,preprint,showpacs]{revtex4}
\usepackage{graphicx}
\usepackage{amsmath,bm}
\usepackage{pstricks}

\begin{document}
\title{Crystal engineering using functionalized adamantane}

\author{J. C. Garcia$^{1}$, L. V. C. Assali$^{1}$, W. V. M. Machado$^{1}$,
and J. F. Justo$^{2}$}

\affiliation{$^1$ Instituto de F\'{\i}sica, Universidade de S\~ao Paulo,\\
CP 66318, CEP 05315-970, S\~ao Paulo, SP, Brazil \\
$^2$ Escola Polit\'ecnica, Universidade de S\~ao Paulo, \\
CP 61548, CEP 05424-970, S\~ao Paulo, SP, Brazil}

\date{June 25, 2010}

\begin{abstract}
We performed a first principles investigation on the structural, electronic,
and optical properties of crystals made of chemically functionalized
adamantane molecules. Several molecular building blocks, formed by boron and
nitrogen substitutional functionalizations, were considered to build
zincblende and wurtzite crystals, and the resulting structures
presented large bulk moduli and cohesive energies, wide and direct bandgaps,
and low dielectric constants (low-$\kappa$ materials). Those properties provide
stability for such structures up to room temperature, superior to those of
typical molecular crystals. This indicates a possible road map for
crystal engineering using functionalized diamondoids, with potential applications
ranging from space filling between conducting wires in nanodevices
to nano-electro-mechanical systems.
\end{abstract}

\maketitle

\section{Introduction}
\label{sec1}

The recent progresses in nanotechnology have allowed to envision
building crystals with designed properties from bottom-up.
This procedure, generally labeled as crystal engineering, could become a
reality within the current available technologies if a number of conditions
are properly observed \cite{wang1,sharma}. The molecular building blocks (MBBs),
the fundamental nanobricks, must carry strong intra-molecular bonding in order
to guarantee internal molecular rigidity \cite{mbb}. Moreover, the MBBs must
contain appropriate chemical active centers to provide attractive and strong
inter-molecular bonding, which ensures rigidity for the entire crystal.
Although getting MBBs with those properties might be reasonably easy to achieve,
the major challenge has been the synthesis, the process in which
those MBBs are put together in organized and systematic fashion. There are
essentially two ways to arrange them: physics- and chemistry-oriented methods.
Physics methods, such as the atomic force microscopy, are used to manipulate
single molecules onto surfaces, allowing positional control with
sub-nanometric precision. However, those methods are fundamentally too slow
because the molecules are manipulated individually and in sequence. On the
other hand, chemistry methods, such as self-assembly, are more appealing
to build nanocrystals in large scale.

The appropriate choice of MBBs is crucial in crystal engineering.
Carbon-based nanostructures, such as nanotubes, buckballs, and
polymers, have been considered as leading candidates to work as
MBBs \cite{moulton}. Diamondoids, also known as molecular diamond,
have been recently included in this list of
candidates \cite{mbb,sasagawa1,garcia,garcia2,mcintosh,drumm2,marsusi}.
Although diamondoids have been known for a long time, they have received
great attention only after recent developments that allowed to manipulate
and separate higher diamondoids \cite{dahl}.

Diamondoids face a major challenge to be used in crystal engineering,
since molecular crystals made of pristine diamondoids have weak inter-molecular
bonding and, consequently, are very brittle \cite{sasagawa1}. An additional
challenge is to find a microscopic driving force to favor self-assembly.
Chemical functionalization is a process in which those two points could
be simultaneously addressed. For example, functionalized diamondoids
could work as MBBs that can self-organize, using the inter-molecular
interactions between chemically active sites in different functionalized
molecules as the driving forces \cite{mbb}. The resulting crystals could
have strong inter-molecular bonds when compared to hydrogen-hydrogen
interactions, that control inter-molecular bonding in typical molecular
crystals.

In order to explore the concept of using functionalized diamondoids as MBBs,
we used first principles calculations to devise crystals
made of functionalized adamantane molecules with boron
and/or nitrogen atoms. We found that the resulting crystals
carried large cohesive energies and bulk moduli, which was a direct consequence
of functionalization \cite{sasagawa1}. Additionally, those crystals present
wide bandgaps and low dielectric constants, equivalent to properties of molecular
crystals made of pristine adamantane \cite{sasagawa1}. All those properties
allow to envision widespread applications, ranging from nano-electro-mechanical
systems (NEMS) to insulating dielectrics in nanodevice interconnects or
opto-electronic devices. This manuscript is organized as follows: section
\ref{sec2} presents the methodological aspects of the simulations,
section \ref{sec3} presents the structural, electronic and optical
properties of the molecular crystals, discussing the role
of functionalization. Finally, section \ref{sec4} discusses
the potential applications for those crystalline structures.

\section{Methodology}
\label{sec2}

The calculations were performed using the VASP package (Vienna {\it ab initio}
simulation package) \cite{kresse1} with periodic boundary conditions. The
electronic exchange-correlation potential was
described within the density functional theory in the generalized gradient
approximation \cite{pbe}. The electronic wave-functions were described by
a projector augmented wave (PAW) method \cite{kresse2}, taking a plane-wave basis
set with an energy cutoff of 450 eV. Convergence in total energy was set to
0.1 meV/atom between two self-consistent iterations. For a crystal
in a certain lattice parameter, configurational optimization was performed
by considering relaxation in all atoms, without any symmetry constrain, until
forces were lower than 3 meV/\AA \ in any atom. The Brillouin zone was sampled
by a Monkhorst-Pack grid of about 100 k-points \cite{mp}. The complex
dielectric tensors were computed within the framework of the random phase
approximation \cite{gajdos}. The dielectric constant ($\kappa$) was computed
as the photon zero-energy limit of the real part of the dielectric tensor.

The reliability of our results using this solid-state-based  methodology was checked
by comparing the properties of isolated adamantane molecules, in pristine and
functionalized forms, with the ones using a chemistry-based methodology
\cite{gaussian,details}. In terms of structural properties of all molecules,
both methodologies provide results for interatomic distances (differences
lower than 0.01 \AA) and angles (differences lower than 0.1$^0$) in
excellent agreement with each other.

\section{Results}
\label{sec3}

\begin{figure}
\centering
\includegraphics[width=13cm]{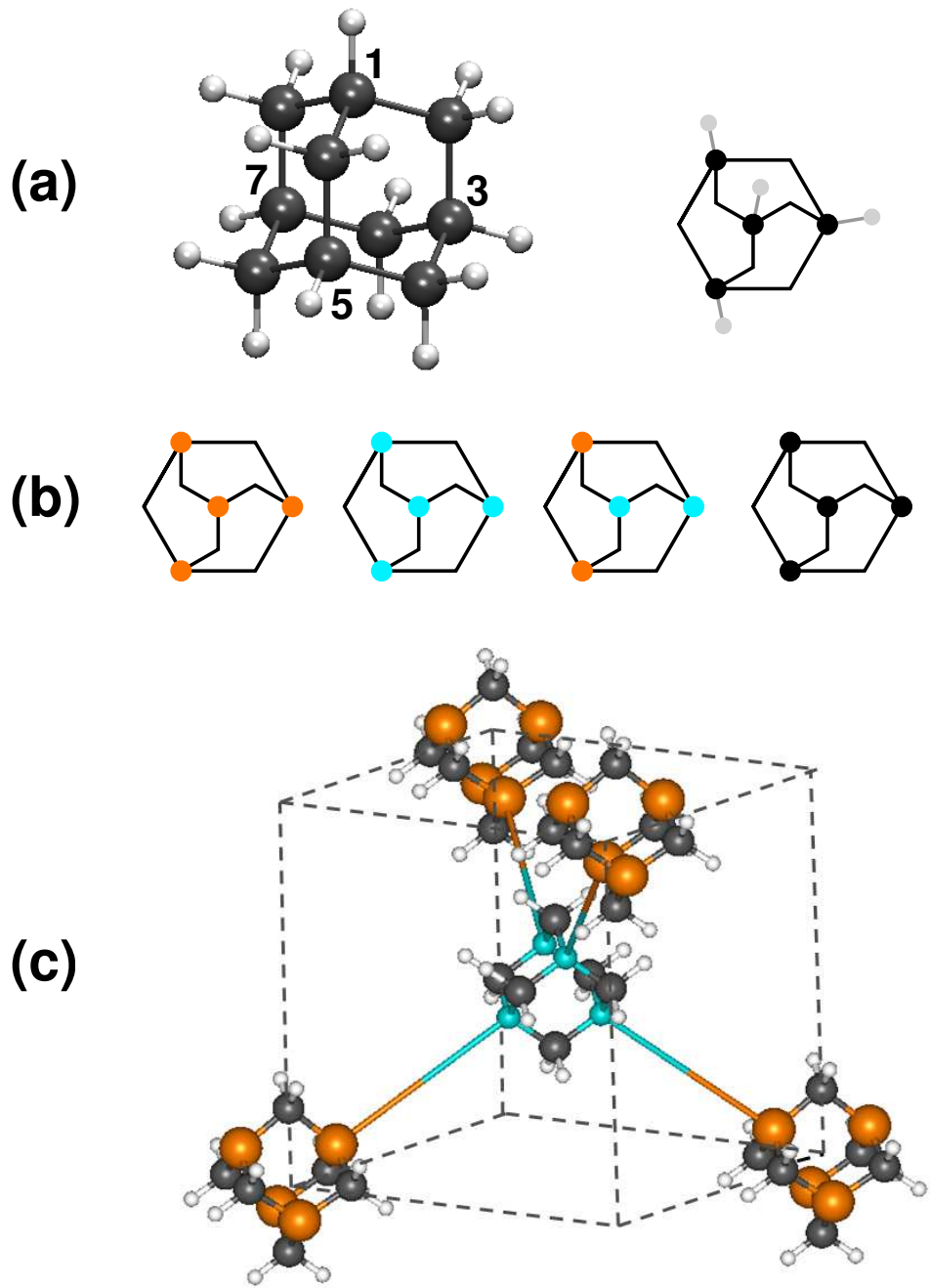}
\caption{(Color online) (a) Atomic configuration and schematic representation
of pure adamantane molecule. The C(1) atoms are labeled by the numbers
1, 3, 5, and 7. The schematic representation shows only the C(1) and their
respective hydrogen atoms. (b) Functionalized molecules: tetra-aza-adamantane
(1,3,5,7-aza-adamantane), tetra-bora-adamantane (1,3,5,7-bora-adamantane),
di-aza-di-bora-adamantane (1,3-aza-5,7-bora-adamantane), and tetra-adamantyl
(1,3,5,7-adamantyl). (c) Configuration of a tetra-aza-adamantane molecule in
a tetrahedral environment of four tetra-bora-adamantane neighbors. Black,
blue, orange, and gray circles represent respectively carbon, nitrogen,
boron, and hydrogen atoms.}
\label{fig1}
\end{figure}

Adamantane, the smallest diamondoid, has a C$_{10}$H$_{16}$ chemical composition,
which comprises of one diamond-like carbon cage with hydrogen atoms passivating
the remaining bonds. It contains two types of carbon atoms: four C(1) and six
C(2) atoms. A C(1) is bound to three C(2)'s  and one H atom, while a C(2) is
bound to two C(1)'s and two H's. Figure \ref{fig1}a presents the structure and
schematic representation of an adamantane molecule. Functionalization in the
C(1) or C(2) sites provides respectively four or six chemically active sites
to allow inter-molecular bonding. In principle, a larger number of
functionalized sites, such as the C(2) ones, would be desirable, since it
could lead to a larger number of inter-molecular bonds, and consequently
stiffer nanostructures \cite{mbb}. However, there are strong evidences that
functionalization with boron or nitrogen atoms is favorable in the
C(1) sites \cite{kamp}.

Chemical functionalization, by replacing the C(1)-H groups with nitrogen and/or boron
atoms, leads to interesting molecules to be used as MBBs:
tetra-aza-adamantane (N$_{4}$C$_{6}$H$_{12}$), tetra-bora-adamantane
(B$_{4}$C$_{6}$H$_{12}$), and di-aza-di-bora-adamantane
(B$_{2}$N$_2$C$_{6}$H$_{12}$), which are labeled here TA, TB and DADB,
respectively, and are schematically represented in figure \ref{fig1}b. Those
molecules are very stable, providing four chemically active sites,
still preserving the strong intra-molecular bonding of the original
adamantane molecule \cite{garcia}. Considering those molecules as MBBs,
self-assembly could take place by using, as the driving force, the dipolar
interaction between the boron site in one molecule and the
nitrogen site in the neighboring one.

In order to explore the potentialities of crystal engineering using
functionalized adamantane, we devised a number of molecular
crystals that could be grown by self-assembly. We should emphasize that
although we are here using the term molecular crystal, the inter-molecular
interactions in those crystals are controlled by covalent bonding \cite{garcia},
instead of the weak dispersive attractive interactions usually observed in
typical molecular crystals. The functionalized adamantane molecules,
with four chemically active sites, have a tetrahedral symmetry, which favors
crystalline organization in zincblende (ZB) or wurtzite (WZ)
structures. The tetrahedral symmetry of such arrangement is indicated in
figure \ref{fig1}c, evidencing that inter-molecular bonding is controlled
by the  B-N interactions. For example, the crystal formed by a basis
of TA and TB molecules in a ZB configuration, labeled here as ZB-TA+TB,
a TA molecule sits in the lattice origin (0,0,0) and a TB one sits in the
(1/4,1/4,1/4)$a$ position, where $a$ is the crystalline lattice parameter.
This crystal could be equivalently formed by a basis consisting
of two DADB molecules, appropriately arranged to provide only B-N
inter-molecular bonds. In this last case, due to the equivalence
of the two molecules in the basis, we labeled as diamond-2DADB.

Crystalline structures with the same molecules discussed in the
previous paragraph, but arranged in a WZ configuration, are labeled WZ-TA+TB
and WZ-2DADB. However, the WZ structure contains four molecules in the
crystalline basis, in contrast to only two in ZB structure. This WZ
structure could be described by a hexagonal structure plus a basis of four
molecules. For example, the WZ-TA+TB is formed by two TA molecules sitting
in (0,0,0) and ($a$/3,2$a$/3,$c$/2) positions and two TB molecules sitting in
(0,0,$uc$) and ($a$/3,2$a$/3,$uc$+$c$/2) positions, where $a$ and $c$
are lattice parameters and $u$ is an internal parameter.

Although the TA+TB and 2DADB crystals, in both crystalline configurations,
are isomeric, they carry important differences in terms of self-assembly.
The TA+TB crystals could be grown layer by layer, with
sub-nanometric precision, by introducing the crystal seed alternately in
solutions rich in tetra-aza- and  tetra-bora-adamantane molecules.
In this case, the driving force for self-assembly comes from attractive Coulomb
interactions between the different molecules in solution.
An equivalent procedure to build 2DADB crystals would be considerably more
difficult using solutions rich in di-aza-di-bora-adamantane molecules,
since all molecules are in the same charge state, and the driving force
should be controlled by the weak attractive multipole interactions.

\begin{figure}
\centering
\includegraphics[width=13cm, angle=0.0]{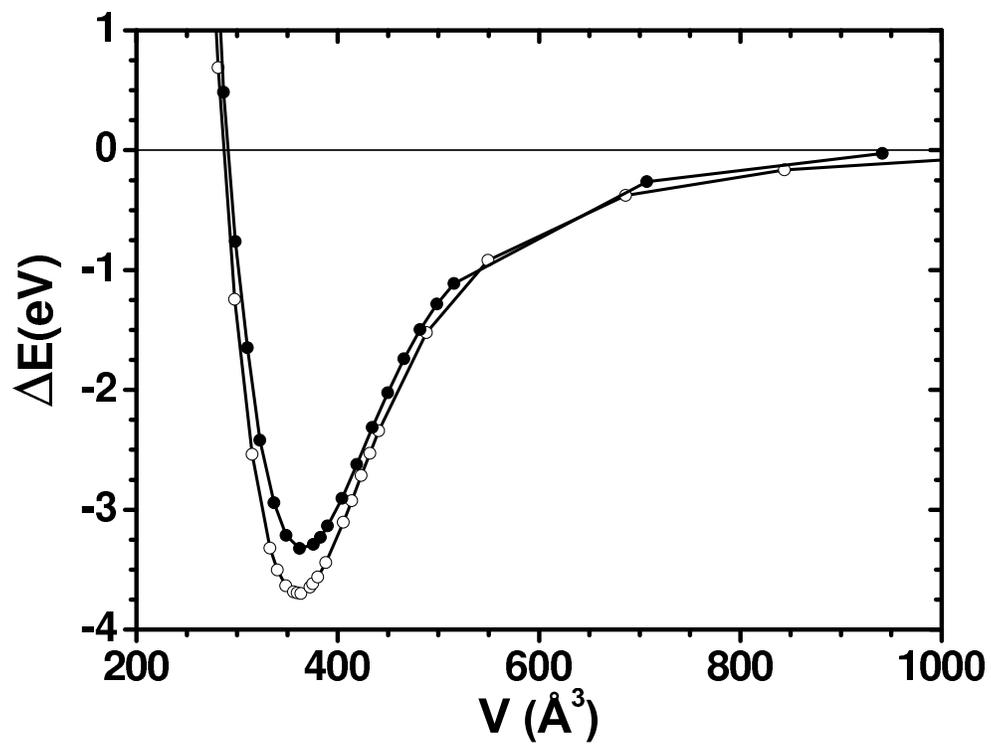}
\caption{Total energy variation per pair of molecules ($\Delta$E),
as a function of their crystalline volumes (V), for the
diamond-2DADB (open symbols) and WZ-2DADB (black symbols) crystals.
The reference energy  is the one for infinitely separated molecules.}
\label{fig2}
\end{figure}

The stability of the molecular crystals can be quantified by the cohesive
(or lattice) energy, the energy gain between a configuration in which the
molecules are infinitely separated and one in a crystalline phase.
Figure \ref{fig2} presents the total energy variation as a function of the
crystalline volume for the ZB- and WZ-2DADB. As discussed in the previous
paragraphs, the molecular crystal in a ZB configuration contains two
functionalized molecules (one TA plus one TB molecule) while the WZ one
contains four molecules (two TA plus two TB molecules), and therefore the
total energy is presented per pair of functionalized molecules,
in order to allow a proper comparison of energies between the ZB and WZ
crystals.
The results indicated strong cohesion for both crystalline structures,
although with a small favoring for the ZB configuration over the WZ one.
Table \ref{tab1} summarizes the properties of crystals with different
compositions, indicating favoring of the ZB structure for all those crystals.

It has been recently shown that cohesion, in molecular crystals made of
functionalized adamantane with boron and nitrogen atoms, is essentially
controlled by the strong inter-molecular
B-N bonding, and intra-molecular bonds are only slightly affected by the
presence of the neighboring functionalized molecules \cite{garcia}.
Therefore, cohesive energy should be essentially associated with the
binding energy of the inter-molecular B-N bonds. For the diamond-2DADB crystal,
cohesive energy per inter-molecular B-N bond is 1.85 eV, which is considerably
smaller than the one for a B-N bond in hexagonal boron nitride solids
of $\approx$ 6.9 eV, computed using the same theoretical framework. However,
this value is large enough to guarantee stability for the molecular crystal.
Additionally, those crystals present large bulk moduli, in the 20-42 GPa range,
when compared to values of around 10 GPa for typical organic molecular
crystals \cite{day}. The 2DADB crystals provide larger cohesive energies
and bulk moduli than the TA+TB ones. The stiffer 2DADB crystals resulted
from the smaller B-N inter-molecular bonds (1.725-1.752 \AA) when compared
to the respective ones (1.907-1.983 \AA) in the TA+TB crystals.
The inter-molecular B-N bonds have a dative covalent character in all crystals,
such that the B-N bonds are
essentially formed by the electrons coming from the non-bonding 2p orbitals
(lone pair) in nitrogen functionalized adamantane.
This covalent behavior was already observed in the interaction between
two functionalized diamantane molecules (one with nitrogen and the other
with boron), as discussed in Ref. \cite{mcintosh}.
This strong inter-molecular
bonding contrasts with bonding in typical molecular crystals, such as crystals
made of pristine adamantane \cite{sasagawa1}, in which bonding is
controlled by the weak dispersive attractive potentials.

We also explored the possibility of building molecular crystals using the
radical tetra-adamantyl (RTA) as a MBB. This molecule is formed by removing
the four hydrogen atoms bonded to C(1) atoms in adamantane
(Fig. \ref{fig1}b). Although this is only a hypothetical molecular
configuration, our calculations indicated  that it is mechanically stable and
maintains the cage-like structure of the original adamantane molecule.
Table \ref{tab1} presents the properties of crystals made of tetra-adamantyl
molecules, called ZB- or WZ-2RTA, indicating large cohesive energies and
bulk moduli. The inter-molecular C-C bonds (1.608-1.632 \AA) in both
configurations are only a little longer than the respective ones in
diamond (1.54 \AA). The binding energy per inter-molecular bond is around
6 eV, which is very close to the cohesive energy per C-C bond in diamond
(7.5 eV). This indicates that the inter-molecular bonds diamond-2RTA and WZ-2RTA
are essentially covalent and equivalent to bonds in diamond.

Our results have so far shown that crystals, made of functionalized
adamantane, have appropriate mechanical properties for several applications
in NEMS. In order to evaluate other potentialities, we explored their
electronic and dielectric properties. All molecular crystals investigated
here have wide and direct bandgaps (between 3.8 and 4.2 eV), as shown in
table \ref{tab1}. These results suggest potential applications for
opto-electronic devices in the ultraviolet region. Those values are about
1 eV lower than the bandgaps computed for molecular crystals made of
pristine adamantane (5.1 eV) using the same methodology. These results are fully
consistent with recent experimental data, which found an 1 eV reduction
in the principal gap of adamantane as result of nitrogen functionalization
(to form tetra-aza-adamantane molecules) \cite{landt1}. As it has been
recently pointed out, the nitrogen atoms incorporated in the C(1) sites of
adamantane introduce 2p nitrogen-related
energy levels in the top of valence band of adamantane,
reducing the principal gap by 1.3 eV \cite{garcia}.

\begin{figure}[h!]
\centering
\includegraphics[width=17cm,angle=0.0]{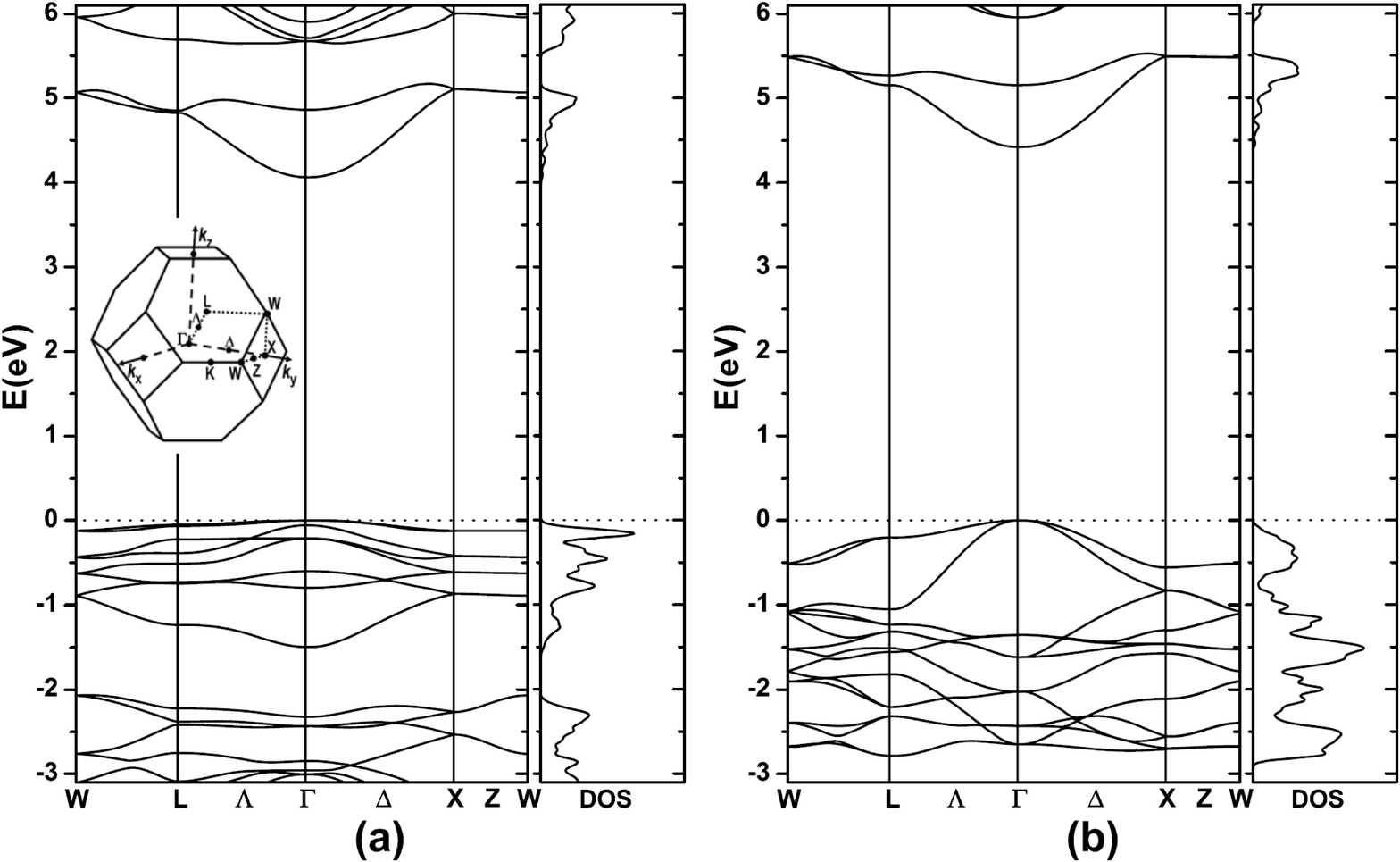}
\caption{Band structure, around the gap region, along several high symmetry
directions and density of states (DOS) of the (a) diamond-2DADB and (b) diamond-2RTA
crystals. The inset in (a) shows the first Brillouin zone and the respective
high symmetry points. The bandgap is defined as the difference in energy
between highest occupied (dashed lines) and lowest unoccupied states.}
\label{fig3}
\end{figure}

Figure \ref{fig3} shows the electronic band structure of the diamond-2DADB
and diamond-2RTA crystals. The band structure of the ZB-TA+TB crystal
\cite{garcia} is similar to that of the diamond-2DADB one, in terms of
dispersion, effective masses and gaps.
In the diamond-2DADB crystal,
the top of the valence band presents a small dispersion, while  the
bottom of the conduction band presents a considerably larger dispersion,
which is totally similar to the band structure of pure adamantane crystal
\cite{sasagawa1}. On the other hand, the diamond-2RTA crystal presents
large dispersion in both band limits, evidencing the
strong covalent bonding controlling inter-molecular interactions.
The effective masses for electrons in
the conduction band and holes in the valence bands
are presented in table \ref{tab1}. The results indicate that
effective masses for holes ($m^{\ast}_{h}$) are
larger than masses for electrons ($m^{\ast}_{e}$) for all
crystals, indicating favoring for electron transport over hole one.

\begin{figure}[h!]
\centering
\includegraphics[width=13.2cm,angle=0.0]{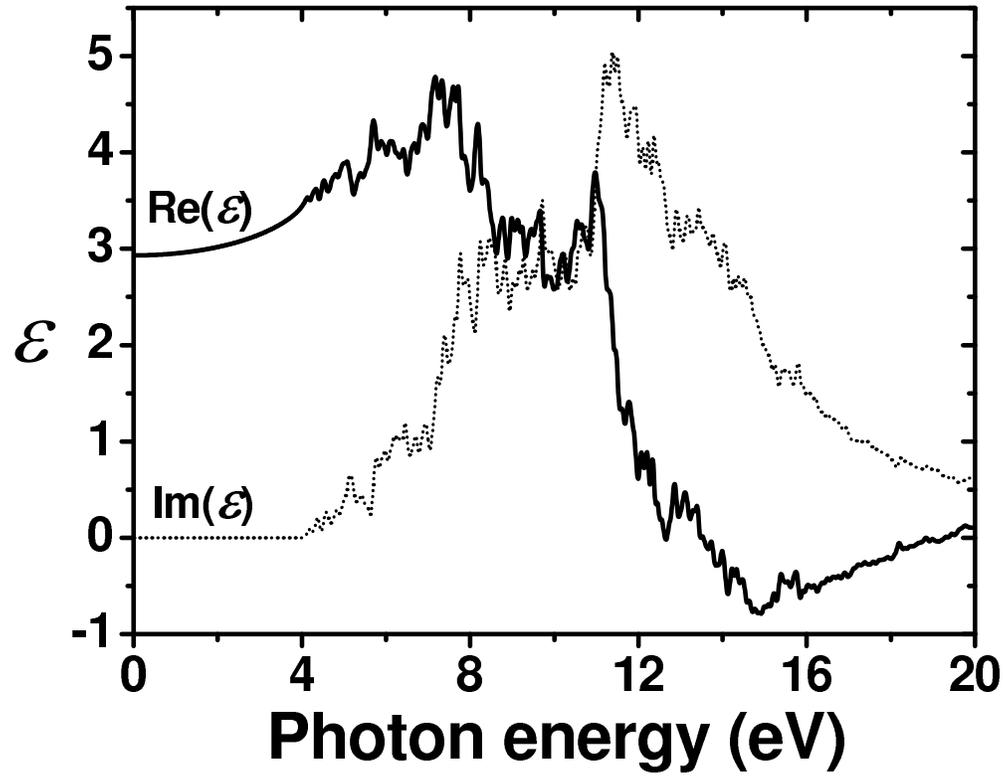}
\caption{Real and imaginary parts of the dielectric function ($\varepsilon$),
versus the photon energy, for the diamond-2DADB crystal. The dielectric constant is
the limit of zero-photon energy of $\rm Re (\varepsilon)$.}
\label{fig4}
\end{figure}

Figure \ref{fig4} shows the real and imaginary dielectric functions
($\varepsilon$) for the diamond-2DADB molecular crystal as a function of the
photon energy. Those functions are similar in all the other molecular
crystals investigated here (TA+TB and 2RTA). The dielectric constant
($\kappa$), computed as the photon zero-energy limit of
$\rm Re (\varepsilon)$, is in the 2.8-3.0 range for all molecular crystals,
which is in the same range of that constant for a molecular crystal made
of pristine adamantane (2.7) in a tetragonal configuration, computed using
the same methodology. Those values are fully consistent with other
calculations on molecular crystals made of pristine diamondoids
(in the 2.46-2.68 range) \cite{sasagawa2}.
Therefore, these molecular crystals could be
classified as low-$\kappa$ materials, in the context of electronic
circuits that consider the dielectric constant of silicon oxide
($\kappa = 3.9$) as a reference value. These results suggest potential
applications as dielectric materials to work as space filling between
nanodevice interconnects to minimize parasitic capacitances.

\section{Summary}
\label{sec4}

In summary, we have shown that adamantane molecules, either chemically
functionalized or in the radical form, are potential candidates to work as
molecular building blocks for nanostructure self-assembly. The
crystals formed by those molecules, in zincblende and wurtzite configurations,
presented favorable mechanical properties, such as large bulk moduli as large
as 70 GPa, which make them very stable at room temperatures or even higher.
All those molecular crystals presented wide and direct electronic bandgaps in the
3.8 to 4.2 eV range, indicating potential applications in opto-electronics
in the ultraviolet region. However, those theoretical gap values should be
considered as lower limits, since they are generally underestimated, when
compared to experimental data, by at least 1 eV using the theoretical
framework of the density functional theory \cite{garcia,mcintosh}.

Those molecular crystals could be used as low-$\kappa$ dielectrics between
interconnects in nanodevices, with dielectric constants in the 2.8-3.0 range.
In the search for low-$\kappa$ dielectrics, scientists have already
found materials in which that constant could be as low as 2.0, such as porous
silicon and silicon oxide materials. Porous materials tend to be
very brittle, and therefore unsuitable for large scale integration in
device technologies. On the other hand, the molecular crystals presented
several advantages over those low-$\kappa$ dielectric materials, since they
provide simultaneously large bulk moduli, which guarantees mechanical
stability.

\vspace{0.5cm}

\textbf{Acknowledgments}

This work was partially supported by the Brazilian Agencies CNPq and CAPES.
The calculations were performed at the computational facilities of
CENAPAD-S\~ao Paulo.


\newpage

\begin{table}[ht]
\caption{Properties of molecular crystals in zinc-blende (ZB), wurtzite (WZ),
  and  diamond configurations,
using several building blocks: tetra-aza-adamantane plus tetra-bora-adamantane
molecules (TA+TB), two di-aza-di-bora-adamantane molecules (2DADB), and
two radical tetra-adamantyl molecules (2RTA). The table presents the
interatomic distance of the inter-molecular bond ($d$), lattice ($a,c$)
and internal ($u$) parameters, cohesive energy per pair of molecules
(E$_c$), bulk modulus (B) and its pressure derivative (B'), electronic
bandgap ($\epsilon_g$), valence band width (VBW),
dielectric constant ($\kappa$), and
hole ($m^{\ast}_{h}$) and electron ($m^{\ast}_{e}$) effective masses,
given in terms of the free electron mass ($m_0$).}
\begin{center}
\begin{tabular}{lcccccc}
\hline \hline
                &  ZB-TA+TB & WZ-TA+TB &  diamond-2DADB &  WZ-2DADB &
                diamond-2RTA & WZ-2RTA \\
\hline
$d$ (\AA)       &  1.907    &  1.983   &   1.725   &   1.752   &  1.608 &  1.632
 \\
$a$ (\AA)       & 11.45     &  8.13    &    11.29  &   8.02    &  11.00 &   7.81
\\
$c$ (\AA)       & ---       &  13.32   &   ---     &   13.11    &  ---&  12.73
\\
$u$          & ---       & 0.377    &  ---      &   0.378    &   ---& 0.382 \\
E$_c$ (eV)                 & 1.81      & 1.67     & 3.70      &  3.32      &
12.11&11.76 \\
B (GPa)                    & 20        & 16  &  43   &  42   & 69   &  72 \\
B'                    & 10.9  & 10.8   & 7.3  &  7.1   &  5.1  & 5.4 \\
$\epsilon_g$ (eV)          & 3.9   & 3.8    & 4.1  &  4.2  & 4.4   &  4.2  \\
VBW (eV)              & 21.3 & 21.3  & 18.9 & 19.1  & 16.6  &  16.7 \\
$\kappa$              & 2.9  &  2.8   & 2.9  &  2.8  & 3.0  & 2.9  \\
$\rm m_e^{\ast}/m_0$   & 0.49 & 0.53   & 0.52 &  0.57 & 0.51 &  0.56   \\
$\rm m_{h}^{\ast}/m_0$ & 2.7  & 3.2    &  2.7 &  3.4 & 1.3  &  1.2   \\
 \hline \hline
\end{tabular}
\end{center}
\label{tab1}
\end{table}

\pagebreak

\end{document}